\documentclass[sigconf]{acmart}

\AtBeginDocument{%
  }

\setcopyright{acmlicensed}
\copyrightyear{2026}
\acmYear{2026}
\acmDOI{XXXXXXX.XXXXXXX}
\acmConference[RecSys '26]{Twentieth ACM Conference on Recommender Systems}{September 27-- October 1, 2026}{Minneapolis, USA}
\acmISBN{978-1-4503-XXXX-X/2018/06}

\usepackage{tcolorbox}
\usepackage{float}
\usepackage[strings]{underscore}
\usepackage[textsize=tiny]{todonotes}

\begin{document}

\title{LLM-Based Re-Ranking for Real Estate Search}


\author{Nkateko Ntimane}
\email{nkateko.ntimane@quintoandar.com.br}
\affiliation{%
  \institution{QuintoAndar}
  \city{Lisbon}
  \country{Portugal}
}

\author{Rafael Guedes}
\email{rafael.guedes@quintoandar.com.br}
\affiliation{%
  \institution{QuintoAndar}
  \city{Lisbon}
  \country{Portugal}
}

\author{Tiago Cunha}
\authornote{Work performed while the author was employed at QuintoAndar.}
\email{tiago.cunha@growthloop.com}
\affiliation{%
  \institution{Growthloop}
  \country{Portugal}
}

\author{Pedro Nogueira}
\email{pedro.nogueira@quintoandar.com.br}
\affiliation{%
  \institution{QuintoAndar}
  \city{Lisbon}
  \country{Portugal}
}

\renewcommand{\shortauthors}{Ntimane et al.}

\begin{abstract}
QuintoAndar Group operates the leading housing marketplace in Latin America for both rentals and sales. 
The platform replaces traditionally paper-heavy workflows with a fully digital experience, making housing transactions faster and more accessible to tenants, buyers, and landlords in the region. Finding the ideal home in such a vast catalog is inherently difficult. At the same time, the widespread adoption of conversational assistants is reshaping user expectations: people increasingly want to express their needs through open, multi-turn dialog rather than rigid filter menus and faceted search. This shift is particularly pronounced in housing, where intent is multi-dimensional, context-dependent, and rarely reducible to a small set of structured constraints. To meet these expectations, we propose a Large Language Model (LLM) based re-ranker that augments a conversational recommendation system by reordering retrieved candidates according to the nuanced, context-rich intent expressed across the user's conversation. We additionally construct a large-scale offline evaluation dataset for conversational real-estate search, containing 960{,}000 query-item pairs constructed from both synthetic and production queries and annotated using an LLM-as-a-Judge framework with human validation. We validate our approach both offline, on this proprietary dataset, and online, through a production A/B test. Both evaluations show consistent improvements in ranking quality, including a statistically significant increase in production of +5.3\% in click-through rate and +4.8\% in scheduled visits, demonstrating the value of integrating conversational context into housing recommendations.
\end{abstract}




\maketitle

\section{Introduction}
\label{sec:introduction}

The widespread adoption of LLM-powered conversational assistants is reshaping how users interact with online services \cite{reuters-ai-news-2026}. Increasingly, people expect to express their needs in natural language, through open and incremental dialog, rather than through rigid forms or filter widgets. This shift has profound implications for recommender systems, particularly in domains where user intent is dense, qualitative, and discovered over time.

Real estate is one such domain. Finding a home is a high-stakes, infrequent decision in which preferences span lifestyle considerations (e.g., proximity to a workplace, sun exposure, neighborhood character), conditional trade-offs (e.g., willingness to exceed a budget for a specific amenity), and qualitative attributes that have no clean representation in a structured catalog. These signals are essentially invisible to traditional faceted search, leading systems to surface listings that satisfy the explicit constraints of a query but miss the user's underlying intent.

At QuintoAndar, this conversational shift has materialized as Concierge, a multi-agent assistant that handles property discovery, listing inquiries, and visit scheduling through natural-language dialog. However, a natural-language interface only results in better recommendations if the ranking stage can take advantage of the additional context provided by the user. Retrieval over structured filters can produce a reasonable candidate-set, but it cannot distinguish between properties that nominally match the same constraints but differ substantially in how well they align with the user's broader goals. Although conversational understanding can be injected at multiple stages of the recommendation pipeline, we focus on the re-ranking stage, which offers a pragmatic balance for production deployments: it operates on a small, latency-tolerant candidate-set, integrates seamlessly with existing high-throughput retrieval infrastructure, and can reason holistically over hard constraints, soft preferences, and relative value.

This work presents the design, deployment, and evaluation of an LLM-based re-ranker integrated into Concierge. The main contributions are as follows:
\begin{enumerate}
    \item An \textbf{LLM-based point-wise re-ranker} that scores candidates using an LLM-based user profile and aggregate statistics over the full candidate-set, enabling relative comparisons under a strict production latency budget.
    \item A \textbf{comprehensive empirical study} combining an offline evaluation on a proprietary dataset of synthetic and production queries, annotated via LLM-as-a-Judge with a production A/B test, both showing consistent gains in ranking quality.
\end{enumerate}

In addition, we describe how we created a large-scale offline benchmark for conversational real-estate search, constructed from both synthetic and real-world production queries. The dataset captures diverse search behaviors, including conversational preferences, trade-offs, lifestyle requirements, and persona-driven constraints, enabling systematic evaluation of ranking models under realistic conditions. 

Our work shows that LLMs can improve search quality in a production recommendation system without requiring end-to-end generative retrieval or a full rewrite of the serving stack. By applying the model only at the re-ranking stage over a compact candidate-set, the approach preserves the latency profile required by interactive search while improving both click-through rate and downstream engagement metrics.

The remainder of the paper is organized as follows. Section~\ref{sec:related} surveys related work on LLM-based recommendation, with a focus on re-ranking and conversational systems. Section~\ref{sec:methodology} describes the proposed user search profile and point-wise LLM re-ranker. Section~\ref{sec:eval} details the experimental setup, including dataset construction, evaluation protocols and results discussion. Section~\ref{sec:conclusion} concludes with a summary of findings and directions for future work.

\section{Related Work}
\label{sec:related}

LLMs have expanded the design space of Recommender Systems (RSs) by enabling natural-language query understanding, richer user-intent modeling, and conversational interaction~\cite{Liu2023}.

\subsection{From Conversational RS to Agentic RS}
\label{subsec:conv_agentic_rs}

Traditional Conversational RSs were defined by multi-turn dialogues where the system explicitly asked questions to elicit user preferences, provide explanations or process the user feedback on the provided recommendations~\cite{Jannach2001}. Earlier approaches focused mainly on critique-based systems, where critiques ("cheaper" or "closer to the center") were used to refine recommendations~\cite{Hammond1994}. Other methods used slot-filling systems that guided users through predefined dialogue paths to acquire specific attributes for recommendation, often constrained by rigid taxonomies~\cite{10.1109/MIS.2007.43}. More recently, neural architectures have been devised to move away from manual dialogue engineering to models trained on large corpora leveraging Reinforcement Learning~\cite{10.1145/3209978.3210002}.

Despite these advances, a gap remains between traditional task-oriented chatbots and systems capable of truly conversational recommendation. This has led to the emergence of Agentic RSs, where the LLM acts as an autonomous reasoning entity, manages the dialogue state implicitly and orchestrates calls to various tools to assist users~\cite{friedman2023leveraging}. 

Recent literature has explored the integration of LLMs into the recommendation pipeline with meaningful success. We categorize this integration into four key stages: 
\begin{itemize}
    \item Query Understanding and Expansion: At the entry point of the pipeline, LLMs serve to bridge the vocabulary mismatch between user natural language and item metadata. By transforming ambiguous queries into structured filters or expanded semantic queries, LLMs improve the precision of initial candidate retrieval~\cite{Nogueira2019, Chuang2023, Ma2023, Yang2025, Liu2025a}. 
    \item User Profiling and Representation: LLMs act as sophisticated feature extractors that distill user history and multi-modal behaviors into compressed, latent representations. This includes generating explicit and detailed user profiles in natural language~\cite{Kim2024,Bao2025,Liu2025} or generating dense embeddings from conversational traces to capture evolving preferences~\cite{Zhu2021, Geng2022, Wang2022, DeNadai2024,xi2026mineandrefine}.
    \item LLM Re-ranking: A two-stage approach where a traditional retriever handles candidate generation, and an LLM acts as a ranker to refine the top-$k$ results. This enables the integration of complex user intent and cross-attribute reasoning that exceeds the capabilities of standard retrieval functions~\cite{Dai2023, Bao2023, Hou2024, Luo2025, Gao2025}.
    \item Agentic Frameworks and Orchestration: A broader form of integration, where the LLM orchestrates the entire process, deciding when to search, when to ask for clarification, and how to present the final recommendation~\cite{Zhang2024, Shi2024, Wang2024, Palumbo2025, Huang2025}.
\end{itemize}

Our work focuses on the re-ranking stage within a conversational real-estate assistant, showing that LLM-based scoring can use conversational context, structured filters, property metadata, and candidate-set statistics to improve ranking quality under production constraints.

\subsection{LLM-Based Re-Ranking}
\label{subsec:llm_reranking}

Using LLMs in the re-ranking stage allows the system to leverage their deep semantic understanding to refine candidate-sets generated by traditional retrievers~\cite{Liu2023}. Unlike first-stage retrieval, which prioritizes efficiency, LLM-based re-ranking focuses on complex reasoning and cross-attribute evaluation.

Early work on LLM re-ranking focused on assessing the zero-shot capabilities of these models, i.e. without explicit training on recommendation data. Research has formalized the task as a conditional ranking problem where the LLM evaluates a list of candidates retrieved by an external model~\cite{Hou2024}. These studies identify three primary paradigms: point-wise, pair-wise, and list-wise ranking~\cite{Dai2023}. While LLMs demonstrate promising alignment with traditional Information Retrieval (IR) ranking, they are often susceptible to the order of items in the prompt or the item's frequency in training data~\cite{Dai2023,Hou2024}.

To overcome the limitations of zero-shot models, recent work emphasizes instruction tuning to align LLMs with specific recommendation objectives. The TALLRec framework demonstrates that efficient tuning can bridge the gap between general language tasks and the recommendation domain~\cite{Bao2023}. More specialized approaches like RecRanker introduce instruction tuning specifically for the top-$k$ list-wise re-ranking task, utilizing importance-aware sampling and position-shifting strategies to eliminate biases~\cite{Luo2025}. The LLM4Rerank framework abstracts different re-ranking objectives (such as accuracy, diversity, and fairness) into a graph-based structure, using Chain-of-Thought (CoT) reasoning to automatically balance these competing factors~\cite{Gao2025}. 

\section{Methodology}
\label{sec:methodology}
This section presents the design of Concierge and its integration with an LLM-based re-ranker. The pipeline follows a two-stage architecture in which a retrieval stage produces a set of candidate properties, and a re-ranking stage refines the candidate set ordering through a point-wise LLM scoring function applied to query-property pairs.

\subsection{Conversational Recommendation System}
Concierge is implemented as a conversational, multi-agent framework that processes natural-language queries and returns personalized property recommendations. The framework comprises several domain-specific agents, including a recommendation agent for property discovery, a property details agent for retrieving listing-specific information, and a visits agent for handling scheduling workflows.

An LLM-based planner acts as the entry point for each user interaction: it classifies the user's intent and dispatches the request to the most appropriate domain-specific agent.

When the planner routes a query to the recommendation agent, the request enters the retrieval pipeline. A core component of this pipeline is Text2Filter, which converts free-text user queries into structured search filters. The supported filter types include location, price range, number of bedrooms, and many others. The service is implemented as a two-stage LLM pipeline based on \textit{GPT-4o (2024-11-20)} \cite{openai2024gpt4o}. In the first stage, an LLM identifies which filter attributes are present in the user query. In the second stage, structured values are generated for each identified attribute. Both stages operate against a predefined schema, ensuring the consistency and validity of the extracted filters.

\begin{figure*}[t]
    \centering
    \includegraphics[width=1\linewidth]{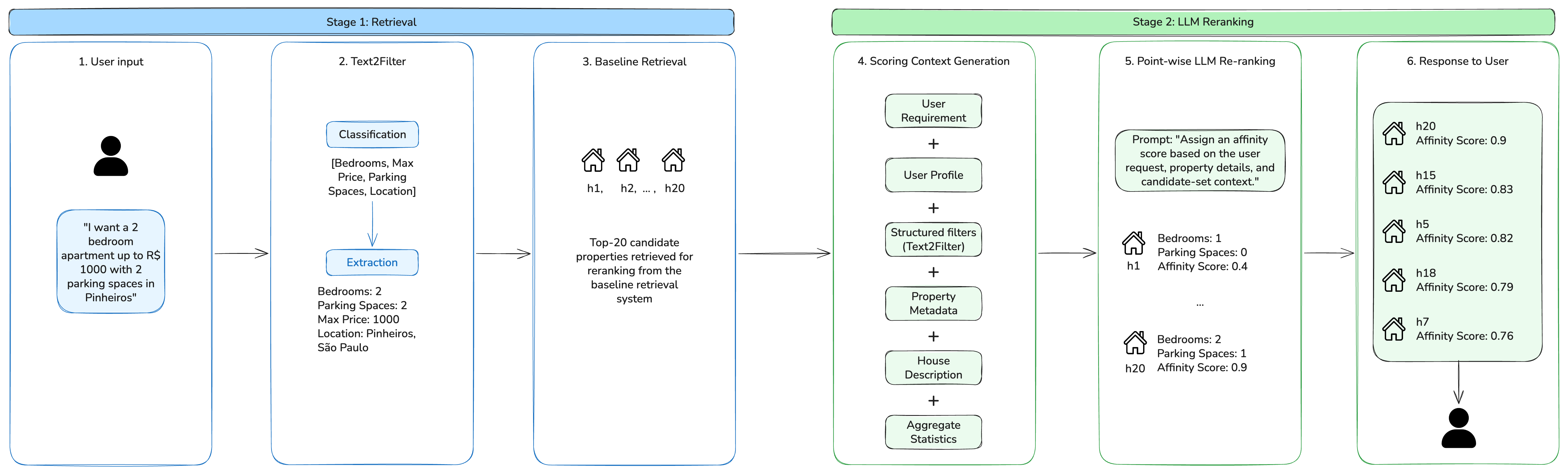}
    \caption{Overview of the LLM re-ranker pipeline.}
    \label{fig:llm_re_ranker_pipeline}
\end{figure*}

The resulting filters are then used to retrieve and rank candidate properties using the existing search system, which serves as the baseline retrieval stage (Figure~\ref{fig:llm_re_ranker_pipeline}). To keep responses concise and easy to interpret in the conversational interface, only the top five ranked properties are returned.
 
\subsection{LLM-based User Search Profile}
\label{subsec:llm_user_search_profile}

The user search profile is a natural-language representation of the user’s search intent. It is constructed by an LLM from three complementary sources: (i) a previously generated profile, when available, providing continuity across sessions; (ii) the most recent conversation with Concierge, capturing explicitly stated requirements; and (iii) a structured behavioral profile derived from historical interactions with listings \cite{zhou2024}. The behavioral profile aggregates implicit preference signals over key attributes, such as location, price, property type, and amenities. It does so by modeling interaction events with different relevance weights according to their importance. For example, clicks are treated as weak engagement signals, whereas favorites and scheduled visits are considered stronger indicators of user preference.

When signals across sources conflict, the model gives precedence to the most recent and most explicitly stated information in a conversation. A central design principle is to treat behavioral signals as soft preferences rather than hard constraints. Instead of converting structured values directly into search filters, the model expresses them as natural-language preferences that capture varying levels of confidence or intensity. This enables the profile to meaningfully distinguish between a weak behavioral tendency and a strong, consistent requirement.

The textual component captures information that is absent from the property catalog. For example, this includes qualitative dimensions such as lifestyle or neighborhood character, as well as conditional trade-offs, such as a user’s willingness to exceed a stated budget for a specific amenity. Such nuanced signals are particularly important in real estate search, where user intent is often multi-dimensional and context-dependent.

The final output is a concise natural-language summary of the user’s search intent together with a list of explicitly mentioned points of interest (Figure~\ref{fig:user_profile_example}). By consolidating implicit behavioral signals and explicit conversational cues into a unified textual representation, the profile furnishes the re-ranker with rich contextual grounding. This enables the re-ranker to surface properties that satisfy persistent or conditional preferences even when those preferences are not directly reflected in the current query.

\begin{figure}[ht]
\centering

\begin{tcolorbox}[
    title=User Search Profile Example,
    colback=gray!5,
    colframe=gray!50,
    width=0.95\linewidth
]
\textit{The user is looking for a house, studio, or apartment to rent in São Paulo, specifically in the South Zone, with a total rental value of less than 1000 reais. Preferred neighborhoods include Brooklin, Butantã, Chácara Santo Antonio, Panamby, and Santo Amaro. They are looking for a property for one person, with at least 1 bedroom and 1 bathroom, and a minimum area of 25 square meters. The user works at Shopping Morumbi and is looking for a property with easy access to the shopping center and near a subway or train station, preferably on the green line. They prefer properties that allow pets, have a glass shower stall, a laundry tub, a fridge, a microwave, kitchen cabinets, are on a silent street, and are faced east.}

\textbf{Points of Interest:} \textit{Shopping Morumbi}
\end{tcolorbox}
\caption{Example of a structured user profile used for personalized real estate retrieval.}
\label{fig:user_profile_example}

\end{figure}

\subsection{LLM-Based Re-ranker}
\label{subsec:llm_baseed_reranker}

 The LLM re-ranker refines the ordering retrieved from the retrieval model to better align with the user’s intent. It does this by applying context-aware scoring using textual and structured inputs together with aggregate statistics computed across the candidate-set. The LLM then produces a refined top-$k$ ranking using a point-wise strategy.

\subsubsection{Point-wise Ranking Strategy}
\label{subsubsec:point-wise}

Under this strategy, each candidate property is scored independently with an affinity score: a scalar estimate of how well the property matches the user’s intent and preferences. This independent scoring allows candidates to be evaluated in parallel, helping keep re-ranking latency low.

In contrast, list-wise methods evaluate the full candidate-set jointly and can capture richer inter-item dependencies, but they are less compatible with our production latency requirements. Pair-wise methods were also considered, but their comparison cost grows quadratically; for example, ranking 20 properties would require 190 pairwise comparisons. We therefore adopt a point-wise formulation as a practical trade-off between ranking quality, scalability, and latency.

\subsubsection{Scoring Context Construction}
\label{subsubsec:scoring_context_constr}

For each request, a scoring context is constructed by combining the following inputs:
\begin{itemize}
    \item \textbf{User requirement:} the free-text request captures the user's primary intent, including explicit constraints, preferences, and qualitative needs that may not be represented in structured fields.
    \item \textbf{Textual user profile:} an LLM-generated summary of historical user interactions provides additional personalization context by capturing recurring behavioral patterns, lifestyle preferences, and implicit property interests that may not be explicitly stated in the current query.
    \item \textbf{Structured filters:} filters derived from Text2Filter provide normalized constraints, such as location, price range, and room requirements, which help the model distinguish hard eligibility criteria from softer preferences.
    \item \textbf{Property metadata:} candidate-level information, including amenities, installations, property attributes, and house descriptions, supplies the evidence used to assess whether each property satisfies the user's stated and inferred preferences.
    \item \textbf{Candidate-set statistics:} aggregate summaries computed over the full candidate-set, such as price ranges, median values, room distributions, property-type distributions, and amenity prevalence, provide contextual baselines for point-wise scoring. Since each property is evaluated independently, these statistics allow the model to judge whether a candidate is comparatively expensive, unusually well equipped, or otherwise distinctive within the retrieved set, improving score consistency across candidates.
\end{itemize}

\subsubsection{Affinity Scoring and Ranking}
\label{subsubsec:affinity_scoring_raknking}

Each candidate is scored by the LLM, which assigns an affinity score between $[0, 1]$ reflecting its alignment with the user’s intent. The scoring prompt considers three complementary dimensions: satisfaction of hard constraints, alignment with stated and inferred user preferences, and relative value within the candidate-set. This formulation combines structured constraint checking with semantic reasoning over trade-offs and implicit preferences. A simplified version of the scoring prompt is shown in Figure~\ref{fig:affinity_score_generation_prompt}. 

Formally, given a user query $q$ and a candidate property $p_i$, the LLM assigns an affinity score:
\[
s_i = f_{\text{LLM}}(q, p_i, u, c)
\]
where $u$ denotes the textual user profile and $c$ represents contextual candidate-set aggregate statistics. The resulting score $s_i \in [0,1]$ estimates how well the property aligns with the user’s intent and preferences.

The final ranked list is then obtained by sorting candidate properties in descending order of affinity score:
\[
R = \text{sort}(P, s)
\]
where $P$ denotes the set of retrieved candidate properties and $R$ denotes the final re-ranked ordering.

Affinity scores are computed asynchronously across all candidates. Properties are then sorted in descending order of affinity score; ties are broken using the original retrieval score produced by the retrieval model. Only the top-$k$ properties are surfaced to the user, where $k=5$ in the production system.

\begin{figure}
\centering

\begin{tcolorbox}[
    title=Affinity Score Generation Prompt,
    colback=gray!5,
    colframe=gray!50,
    width=0.95\linewidth
]

\small

You are an Affinity Score Generator in a real estate recommender system. Your task is to assign an affinity score between 0.0 and 1.0 indicating how well the property matches the user's intent.\\

You are provided the following: \\
(i) the user requirement; \\
(ii) structured user filters; \\
(iii) textual user profile; \\
(iv) metadata and textual description for a single candidate property; and \\
(v) aggregate statistics computed across the candidate-set.\\

Scoring should consider:
\begin{itemize}
    \item alignment with explicit constraints (e.g., location, price, bedrooms, etc.) and user preferences;
    \item semantic information from property descriptions;
    \item value relative to other candidate properties;
    \item rarity and usefulness of amenities and installations; and
    \item consistency of scores across independently evaluated candidates.\\
\end{itemize}

The full scoring range should be used, where higher scores indicate stronger alignment with the user's preferences and contextual requirements.

\end{tcolorbox}

\caption{Simplified affinity score generation prompt used for point-wise LLM-based re-ranking.}
\label{fig:affinity_score_generation_prompt} 

\end{figure}

\section{Evaluation}
\label{sec:eval}

We evaluate the LLM-based re-ranker along two complementary axes. First, offline experiments on a labeled search dataset measure intrinsic ranking quality under controlled conditions, enabling systematic comparison of configurations and ablation studies.
Second, we conducted an online A/B test in production to measure the downstream impact of the re-ranker on real user interactions. In this setting, behavioral signals such as clicks and downstream engagement serve as proxies for user satisfaction. Together, offline and online evaluations provide a comprehensive view of both ranking quality and real-world effectiveness.

\subsection{Offline Search Dataset}
\label{subsec:offline_search_dataset}

To enable robust and scalable evaluation of LLM-based re-ranking strategies, we construct a large-scale, domain-specific offline dataset. Our approach combines synthetic query generation with real-world production queries, and leverages an LLM-as-a-Judge framework for scalable relevance annotation, complemented by human validation. This design approximates realistic search behavior while avoiding the considerable cost of fully manual labeling.

\begin{figure*}[t]
    \centering
    \includegraphics[width=\textwidth]{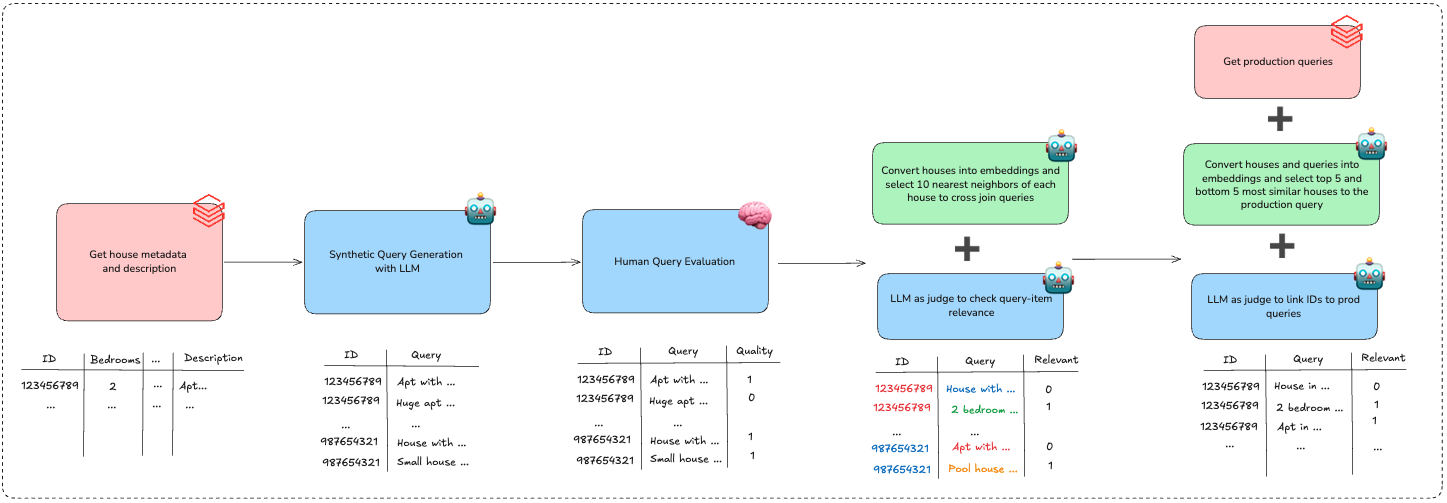}
    \caption{Overview of the dataset construction pipeline. (1) Property data is sampled from the data lake. (2) Candidate listings are retrieved using dense embeddings to construct hard negatives. (3) Relevance labels are assigned using an LLM-as-a-Judge. (4) The dataset is enriched with real-world queries.}
    \label{fig:dataset_pipeline}
\end{figure*}

The dataset construction pipeline consists of four stages (Figure~\ref{fig:dataset_pipeline}).

\subsubsection{Data Selection and Synthetic Query Generation}
\label{subsubsec:synthetic_query_gen}

We randomly sampled 10{,}000 active property listings from QuintoAndar's inventory. Structured fields provide core attribute signals, while descriptions contribute complementary information, such as proximity to points of interest and qualitative neighborhood characteristics.

Synthetic queries are generated with \textit{Gemini 3.1 Flash Lite} \cite{google2026gemini31flashlite} by treating each listing as an anchor item. We define a taxonomy of six query types to capture diverse search behaviors: (i)~explicit attribute queries, (ii)~trade-off formulations, (iii)~high-complexity constraint queries, (iv)~conversational queries, (v)~lifestyle and amenity driven queries, and (vi)~persona-based scenarios~\cite{palumbo2025audioboost}. One query per category is generated for each property, yielding 60{,}000 synthetic queries in total. 
This taxonomy ensures systematic coverage of complex and underrepresented patterns, such as multi-constraint or persona-driven queries, that are sparse in production logs but critical for evaluating advanced retrieval systems.

Query quality is validated through a human evaluation on a random sample of 1,000 query-property pairs, where annotators assign binary relevance labels. We observe 94\% agreement with the LLM-generated labels, which confirms high synthetic fidelity.

\subsubsection{Candidate Retrieval and Hard Negative Generation}
\label{subsubsec:negative_mining}

For each anchor property, we retrieve semantically similar candidates to enable the inclusion of hard negatives rather than trivial random ones. Given the cross-lingual nature of the setting (English structured metadata and Portuguese descriptions), we concatenate both field types and compute dense embeddings using \textit{intfloat/multilingual-e5-base}~\cite{wang2022textembeddings}, which offers a favorable trade-off between retrieval performance and computational efficiency. Embeddings are indexed with FAISS~\cite{douze2024faiss} under an inner-product similarity metric. For each anchor we retrieve the top-$k$ nearest neighbors ($k=10$), expanding the dataset from 60{,}000 to 600{,}000 query–item pairs while keeping LLM-based annotation tractable \cite{xi2026mineandrefine}.

\subsubsection{Relevance Annotation via LLM-as-a-Judge}
\label{subsubsec:relevance_annotation}

Binary relevance labels are assigned using \textit{Claude Sonnet 4}~\cite{anthropic2025claude4} with a structured prompting strategy: the model is instructed to (i)~identify all constraints expressed in the query, (ii)~verify their presence in the candidate listing, and (iii)~apply strict rejection criteria for mismatches on critical attributes such as location and price range. Another human evaluation was performed on a sample of 1000 pairs yielding to 96\%\ agreement with LLM judgments, supporting the use of LLM-as-a-Judge as a scalable annotation alternative.

\subsubsection{Integration of Production Queries}
\label{subsubsec:integrate_prod_queries}

To complement synthetic data with realistic user behavior, we sampled 30{,}000 unique queries from production logs and incorporate them into the dataset. For each query, candidate listings are retrieved using the same embedding model and FAISS configuration from stage 2. To ensure a diverse evaluation set, candidate pools are constructed by combining the top-5 most similar items (likely positives and hard negatives) with the bottom-5 least similar items (easy negatives), capturing a broad spectrum of difficulty levels. Relevance annotation is performed using the same LLM-as-a-Judge framework. 

Combining synthetic and production queries provides complementary benefits: synthetic data increases coverage and complexity, while real queries reflect authentic user intent.

\subsubsection{Dataset Summary}
\label{subsubsec:dataset_summary}

The final dataset contains 960,000 query–item pairs. Each pair consists of a user query and a candidate property listing, represented by its identifier, structured metadata, unstructured description, and a binary relevance label. The relevance distribution is 33\% positive and 67\% negative, reflecting a moderately imbalanced setting that increases the difficulty of creating an accurate ranking.

\subsection{Experimental Variations}
\label{subsec:experimental_variations}

All experimental variants are derived from the affinity-scoring prompt introduced in Section~\ref{subsubsec:affinity_scoring_raknking} and illustrated in Figure~\ref{fig:affinity_score_generation_prompt}.

We first evaluate the baseline retrieval system without LLM-based re-ranking. We then incrementally introduce the proposed re-ranking components to measure their individual contributions to the ranking quality.

The evaluated variations include:
\begin{itemize}
    \item \textbf{Baseline (B):} Existing retrieval system. 
    
    \item \textbf{Base Re-ranker (BR):} Introduces the point-wise LLM re-ranking stage using structured property metadata and user query information.
    
    \item \textbf{Aggregated Statistics (AS):} Adds candidate-set statistics such as price ranges, room distributions, and amenity prevalence to provide contextual grounding for point-wise scoring.
    
    \item \textbf{Scoring Rules (SR):} Adds explicit scoring calibration instructions to encourage globally consistent affinity scores across independently evaluated candidates.
    
    \item \textbf{House Description (HD):} Adds unstructured property descriptions to evaluate the contribution of richer semantic property context.
    
    \item \textbf{Textual User Profile (TUP):} Adds an LLM-generated textual representation of historical user preferences to evaluate the effect of personalization signals.
\end{itemize}

\subsection{Offline Evaluation and Results}
\label{subsec:offline_eval_and_results}
All configurations are evaluated on the dataset described in Section~\ref{subsec:offline_search_dataset}.

\subsubsection{Results}
\label{subsec:offline_results}

In table~\ref{tab:offline_results} we present the offline ranking performance across selected prompt configurations and experimental variations.

\begin{table}
\centering
\caption{Offline ranking performance across prompt formulations and input configurations.}
\label{tab:offline_results}
\resizebox{\linewidth}{!}{
\begin{tabular}{lcccc}
\toprule
\textbf{Configuration} & \textbf{Recall@5} & \textbf{nDCG@5} \\
\midrule
B + BR + AS + HD + TUP & \textbf{0.814*} & \textbf{0.889*} \\
B + BR + AS + HD & \underline{0.797} & \underline{0.865*} \\
B + BR + AS + SR & 0.784 & 0.848* \\
B + BR + AS & 0.790 & 0.857* \\
B + BR & 0.784 & 0.850* \\
B & 0.766 & 0.770 \\

\bottomrule
\end{tabular}
}

\vspace{0.5em}
\footnotesize{* Statistically significant compared to the baseline configuration ($p < 0.05$).}

\end{table}

The LLM-based re-ranker substantially improved ranking quality over the baseline retrieval system across all evaluated metrics. The strongest improvements were observed in nDCG, indicating that the re-ranker is particularly effective at improving the ordering of relevant properties near the top of the ranked list. The base re-ranker improved nDCG@5 by 10.4\% relative to the baseline retrieval system. Furthermore, all evaluated re-ranking configurations achieved statistically significant improvements over the baseline configuration in nDCG@5.

Adding candidate-set aggregate statistics yielded a consistent gain, with a 0.8\% increase in nDCG@5 versus the matched configuration without these statistics. This suggests that candidate-level context helps the model compare relative value, rarity, and trade-offs during point-wise scoring.

In contrast, explicit scoring rules slightly degraded ranking performance, reducing nDCG@5 by 1.1\% relative to the corresponding configuration without scoring calibration instructions. This suggests that rigid score calibration instructions may reduce the model’s ability to make nuanced semantic ranking decisions in the point-wise setting.

\begin{figure}
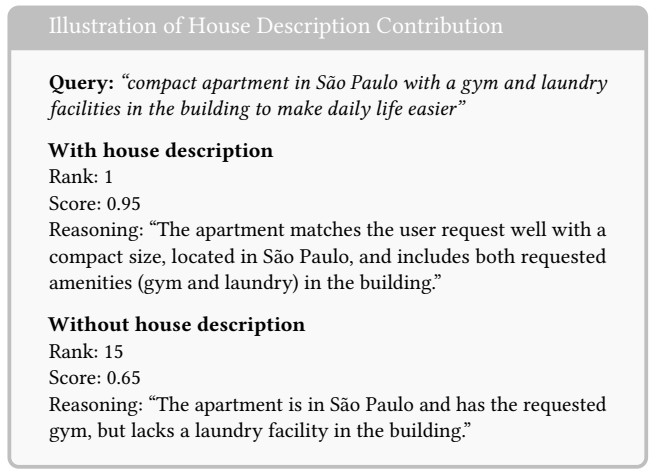

\centering
\begin{tcolorbox}[
    title=Illustration of House Description Contribution,
    colback=gray!5,
    colframe=gray!50,
    width=\linewidth
]

\small

\textbf{Query:}
\textit{“compact apartment in São Paulo with a gym and laundry facilities in the building to make daily life easier”}

\vspace{0.2cm}
\textbf{With house description}

Rank: 1 \\
Score: 0.95\\
Reasoning: “The apartment matches the user request well with a compact size, located in São Paulo, and includes both requested amenities (gym and laundry) in the building.”

\vspace{0.2cm}
\textbf{Without house description}

Rank: 15\\
Score: 0.65\\
Reasoning: “The apartment is in São Paulo and has the requested gym, but lacks a laundry facility in the building.”

\end{tcolorbox}

\caption{Example illustrating the contribution of house descriptions.}
\label{fig:long_description_example}
\end{figure}

House descriptions also contributed positively to the ranking metrics, improving nDCG@5 by 0.9\% relative to the corresponding configuration without descriptions. These results indicate that free-text property descriptions provide important semantic signals beyond structured metadata alone, particularly for lifestyle-oriented and qualitative queries. Figure~\ref{fig:long_description_example} presents an illustrative example of this effect.

Textual user profiles produced the strongest overall performance improvements. The best-performing configuration combined aggregated statistics, house descriptions, and textual user profiles, improving nDCG@5 by 2.8\% and Recall@5 by 2.1\% relative to the strongest configuration without textual user profiles. These results suggest that combining richer semantic property context with historical preference representations provides stronger personalization signals for ranking semantically similar properties.

The second-best configuration, which excludes textual user profiles, corresponds to the version deployed in the online A/B experiment presented in Section~\ref{subsec:online_ab_testing}. At the time of deployment, textual user profiles were available for offline evaluation but had not yet been integrated into the production serving pipeline.

Overall, the offline results show that the proposed LLM-based re-ranker improves both retrieval coverage and ranking quality relative to the baseline retrieval system. This suggests that the re-ranker is particularly effective at improving the ordering of relevant candidate properties near the top of the ranked list. These results further suggest that richer semantic context, including house descriptions, aggregated candidate-set statistics, and textual user profiles, enables the model to make more effective ranking decisions for semantically similar properties. These findings are consistent with the role of the re-ranker as a refinement stage operating over a strong retrieval baseline.

\subsection{Online A/B Testing}
\label{subsec:online_ab_testing}

The re-ranker was deployed in a randomized A/B experiment conducted between April 7 and May 8, 2026. The users were randomly assigned to the treatment and control groups with a 50/50 allocation. The experiment included more than 200,000 production recommendation traces collected from real user interactions with Concierge. The treatment group received LLM re-ranked results, while the control group received the baseline retrieval ordering. Key evaluation metrics included the Click-Through-Rate (CTR) and scheduled visits through the conversation system.

\begin{figure}
    \centering
    \includegraphics[width=\columnwidth]{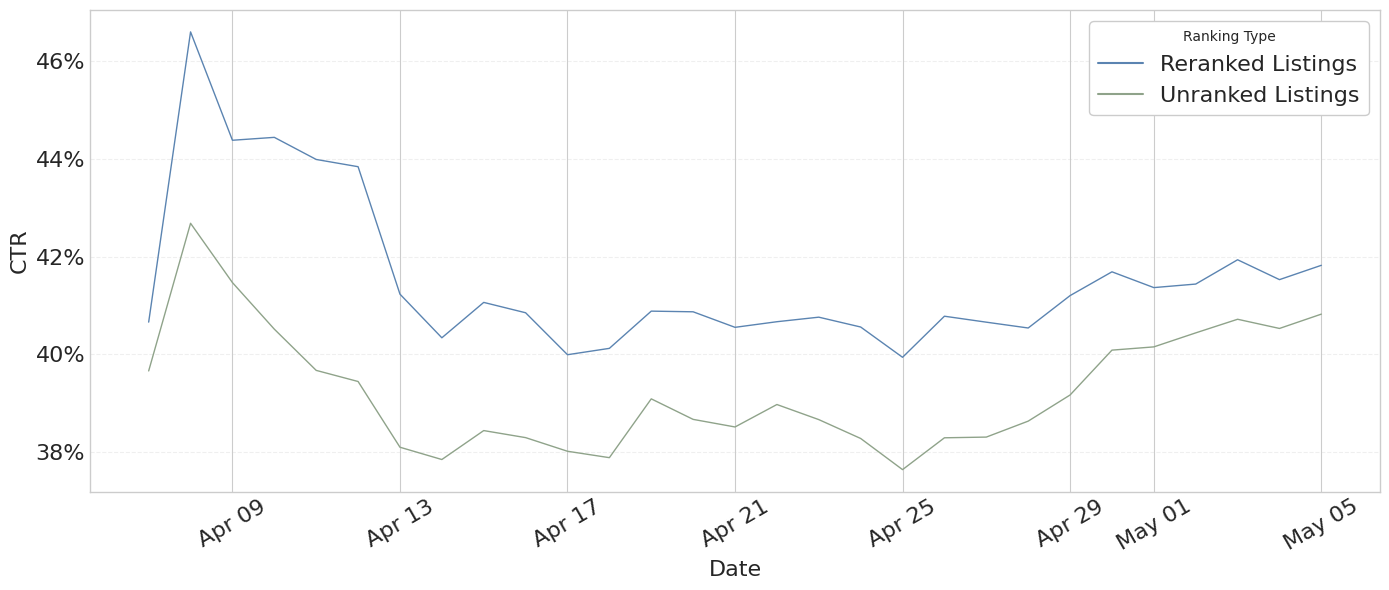}
    \caption{CTR Over Time by Ranking Type during the A/B test.}
    \label{fig:ctr_over_ranking_type}
\end{figure}

In the A/B experiment, the LLM-based re-ranker consistently outperformed the retrieval baseline in CTR throughout the evaluation period (Figure~\ref{fig:ctr_over_ranking_type}), achieving a statistically significant +5.3\% relative CTR improvement. The impact also extended beyond clicks: the treatment produced a statistically significant +4.8\% uplift in scheduled visits generated through the conversational system. Together, these results show that the re-ranker improved both immediate recommendation engagement and downstream real-estate outcomes in production, suggesting that better ranking quality translated into more meaningful user actions rather than only higher surface-level engagement.

From a systems perspective, the online deployment validated the practicality of the proposed point-wise architecture. Despite the additional LLM inference stage, the system maintained stable production latency throughout the experiment and remained within the operational response-time constraints of the conversational platform. Compared with the retrieval baseline, end-to-end response latency increased by an average of 4.2s, while operational inference costs increased by approximately +7\%. Although these overheads represent a measurable deployment cost, they remained within acceptable production limits and were outweighed by the observed improvements in user engagement and scheduled visits.

\subsection{LLM-as-a-Judge Evaluation}
\label{subsec:llm_as_a_judge}

In addition to A/B testing, the re-ranker is evaluated on sampled production traces to assess ranking quality in a controlled setting. A LLM-as-a-Judge model compares the ordering of properties before and after re-ranking. For each trace, the user query and the top five properties from both the retrieval model and re-ranked lists are presented in anonymized form. The judge selects which ordering better satisfies the user intent.

The evaluation is formulated as a pairwise preference task rather than an absolute scoring problem. This design reduces score calibration issues and allows the judge to focus directly on comparative ranking quality. The judge is instructed to prioritize hard constraints such as location, budget, and number of bedrooms, while also considering softer preferences, semantic relevance, and value-for-money trade-offs. To better reflect realistic recommendation behavior, properties that slightly violate constraints may still be preferred when they offer substantially better overall value.

The candidate lists are anonymized and presented without identifying information to reduce positional and presentation bias.

\begin{figure}
    \centering
    \includegraphics[width=\linewidth]{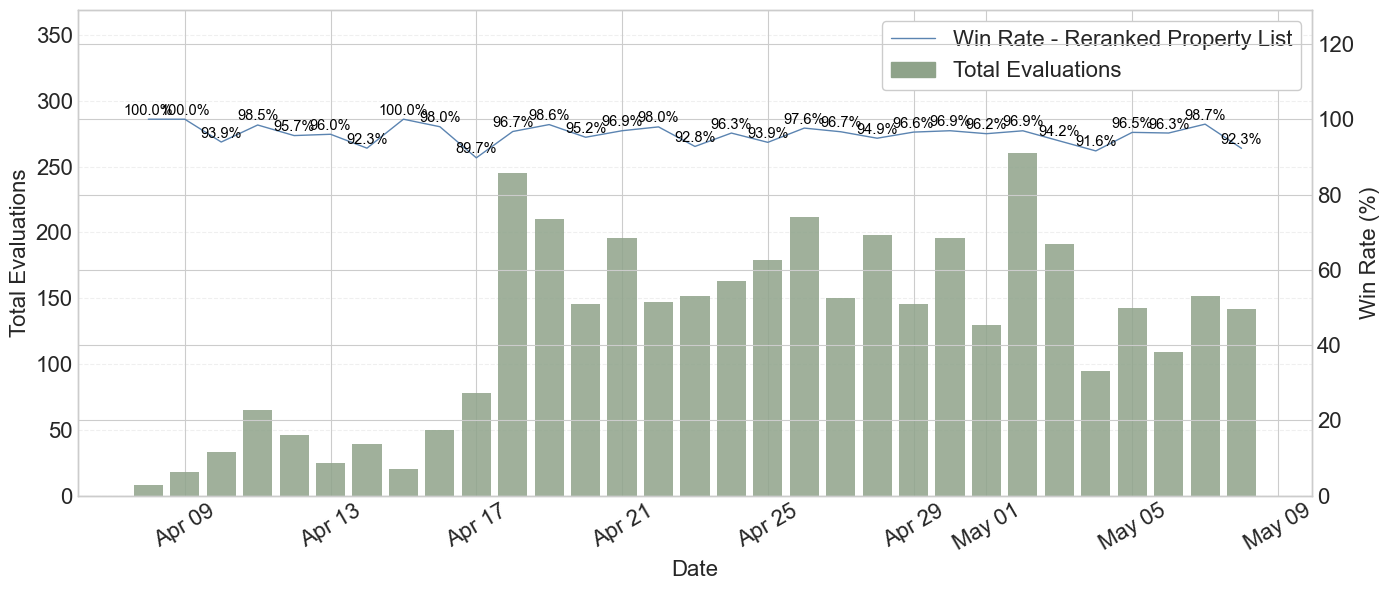}
    \caption{LLM-as-a-Judge win rate and evaluation volume}
    \label{fig:llm_judge_results}
\end{figure}

The LLM-as-a-Judge evaluation further supports the production results. Across 3,944 sampled traces, the judge model preferred the re-ranked property lists over the baseline retrieval ordering in 95\% of evaluated cases. Figure~\ref{fig:llm_judge_results} shows the daily win rate and evaluation volume throughout the evaluation period.
Qualitative inspection of the judge rationales showed that the re-ranker more effectively prioritized:
\begin{itemize}
    \item Properties satisfying explicit user constraints.
    \item Stronger value-for-money trade-offs.
    \item Semantically relevant lifestyle and amenity preferences.
    \item Higher-quality properties earlier in the ranked list.
\end{itemize}

The judge explanations also revealed that the re-ranker frequently improved the ordering of semantically ambiguous or borderline candidates. At the same time, it preserved the strongest matches identified by the retrieval system. These observations are consistent with the improvements in ranking quality observed in the offline metrics results.

\section{Conclusion}
\label{sec:conclusion}

This paper presented an LLM-based point-wise re-ranker for Concierge, a production conversational assistant for real-estate search. By combining conversational context, structured filters, property metadata, house descriptions, and aggregate statistics from the candidate-set, the proposed architecture improves the quality of the ranking while preserving the scalability benefits of parallel candidate scoring. Offline and online evaluations show gains in intrinsic ranking quality, click-through rate, and downstream engagement signals. Importantly, the production deployment maintained stable latency despite the additional LLM inference stage, suggesting that asynchronous point-wise scoring is an operationally feasible approach for LLM-based conversational recommendation systems.

\section{Future Work}
\label{sec:future_work}

Future work will explore deeper agentic integration, where retrieval, ranking, and follow-up interactions are optimized jointly using the broader conversational state. This includes A/B testing the addition of the LLM-based user profile as a personalization signal, allowing the assistant to ask clarifying questions when user intent is ambiguous, adapting the candidate-set dynamically across conversation turns, and incorporating feedback from downstream actions such as clicks, saved properties, and scheduled visits. Another promising direction is to investigate hybrid point-wise and list-wise strategies that preserve the latency advantages of independent scoring while adding richer inter-candidate reasoning when the candidate-set is small or the query requires more nuanced trade-off analysis.

\section{AI Usage Disclosure}

The authors used AI-assisted tools for grammar correction and minor editorial improvements. All scientific content, conclusions, and claims are solely the responsibility of the authors, who vouch for the accuracy, originality, and integrity of the work presented.

\section{Authors} 

Nkateko is a Data Scientist at QuintoAndar, working on conversational search and recommendation systems. Previously, she worked across the fintech, consulting and insurance industries, developing machine learning solutions, recommendation models and large-scale data pipelines. Nkateko is also actively involved in technology mentorship and digital skills initiatives across Africa. She holds a BSc in Applied Statistics \& Economics from the University of Cape Town. \\

Rafael is a Senior Data Scientist at QuintoAndar, working on conversational search and recommendation systems. Previously, he focused on recommendation challenges at a food delivery company. With more than 7 years of experience in the field, Rafael is passionate about applying machine learning to solve real-world business problems and build impactful data-driven solutions. He holds an MSc in Data Science \& Engineering from the University of Porto.  \\

Pedro Nogueira is a Head of Data Science at QuintoAndar working in search and recommendations. Previously he also tackled Information Retrieval problems at a luxury fashion ecommerce company. With more than 10 years experience under his belt, Pedro’s passion is to build machine learning systems that have real impact in companies. He holds an MSc in Electrical and Computer Engineering from University of Porto. \\

Tiago is a Senior Staff AI/ML Researcher at Growthloop, working on Agentic Decisioning and Outcomes-based Marketing Optimizations. He has over 6 years of industry experience building data-driven decision systems in large-scale marketplaces and e-commerce environments. Previously, he worked on search and ranking problems in the travel and real estate domains. He holds a PhD in Computer Science from the University of Porto. \\ 

\bibliographystyle{ACM-Reference-Format}
\bibliography{sample-base}

\end{document}